\documentclass[conference]{IEEEtran}
\usepackage{graphicx}
\usepackage{url}

\ifCLASSINFOpdf
\else
\fi
\usepackage[pscoord]{eso-pic} 

\begin{document}
%
% paper title
% Titles are generally capitalized except for words such as a, an, and, as,
% at, but, by, for, in, nor, of, on, or, the, to and up, which are usually
% not capitalized unless they are the first or last word of the title.
% Linebreaks \\ can be used within to get better formatting as desired.
% Do not put math or special symbols in the title.
\title{On optimizing Inband Telemetry systems for accurate latency-based service deployments}

% author names and affiliations
% use a multiple column layout for up to three different
% affiliations

%\author{Nataliia Koneva, Alfonso S\'{a}nchez-Maci\'{a}n, Jos\'{e} Alberto Hern\'{a}ndez, \'{O}. Gonz\'{a}lez de Dios \\
%Universidad Carlos III de Madrid, Spain \\
%Telefonica I+D, Spain
    % <-this % stops a space
%\thanks{N. Koneva, A. Sanchez-Macian and J. A. Hernandez are with the Dept. Ing. Telematica at Universidad Carlos III de Madrid, Spain.}% <-this % stops a space
%}

\author{
    \IEEEauthorblockN{Nataliia Koneva\IEEEauthorrefmark{1}, Alfonso S\'{a}nchez-Maci\'{a}n\IEEEauthorrefmark{1}, Jos\'{e} Alberto Hern\'{a}ndez\IEEEauthorrefmark{1}, \'{O}. Gonz\'{a}lez de Dios\IEEEauthorrefmark{2}}
    \IEEEauthorblockA{\IEEEauthorrefmark{1}Universidad Carlos III de Madrid, Spain
    \\nkoneva@pa.uc3m.es , \{alfonsan, jahgutie\}@it.uc3m.es}
    \IEEEauthorblockA{\IEEEauthorrefmark{2}Telefonica I+D, Spain
    \\oscar.gonzalezdedios@telefonica.com}
}

\maketitle

% As a general rule, do not put math, special symbols or citations
% in the abstract
\begin{abstract}
The power of Machine Learning and Artificial Intelligence algorithms based on collected datasets, along with the programmability and flexibility provided by Software Defined Networking can provide the building blocks for constructing the so-called Zero-Touch Network and Service Management systems. However, the fuel towards this goal relies on the availability of sufficient and good-quality data collected from measurements and telemetry. This article provides a telemetry methodology to collect accurate latency measurements, as a first step toward building intelligent control planes that make correct decisions based on precise information.
\end{abstract}

% no keywords

% For peer review papers, you can put extra information on the cover
% page as needed:
% \ifCLASSOPTIONpeerreview
% \begin{center} \bfseries EDICS Category: 3-BBND \end{center}
% \fi
%
% For peerreview papers, this IEEEtran command inserts a page break and
% creates the second title. It will be ignored for other modes.
\IEEEpeerreviewmaketitle

\section{Introduction}
% no \IEEEPARstart
Software-Defined Networking (SDN) paves the path to the realization of intelligent networks by enabling comprehensive network automation and zero-touch network and service management (ZSM), encompassing essential aspects of autonomous self-adaptation, self-healing mechanisms, and on-demand re-configuration~\cite{gallego_zsm}. Network automation has been traditionally perceived as a closed-loop process comprising three key tasks: (1) the continuous monitoring and data collection of network states, covering both optical and packet-based aspects; (2) the execution of advanced, intelligent algorithms, including AI/ML for evaluating network performance and optimizing the placement of traffic flows to enhance user quality of experience; (3) the application of recommendations generated by these intelligent algorithms. This is often referred to as the 'observe-decide-react' loop. 

It is then critical to design accurate telemetry systems that collect sufficient and relevant measurements since, if the information collected in step one is not good enough, then the decisions taken in step two may be totally wrong. In this article, we focus on the such step one:  monitoring and collecting information from in-band telemetry, in particular, for estimating end-to-end latency. Specifically, our focus lies in identifying the optimal methodology for gathering measurements in a manner that strikes the balance between obtaining a sufficient number of latency samples to ensure accuracy and avoiding excessive monitoring which could lead to network overload.

\section{Methodology and examples}
\label{sec:background}

Indeed, latency is critical for most existing and emerging Internet applications such as video-on-demand, tele-surgery, and metaverse-related applications~\cite{modyano_ieeeproc,hernandez_rtt}. In particular, Tactile Internet sets 1~ms as the desired maximum end-to-end latency, while virtual reality and metaverse applications are even stricter with 0.1~ms as latency target. Often, some of these services have to traverse the access, aggregation and metro network segments, heading toward a datacenter tens of kilometers away, where such services can be processed.%~\cite{JoseAlbertoCloudification}. 

%One of the parameters that have an important impact on the quality of service of applications that make use of the network is the delay. The network delay is formed by different components such as transmission delay, propagation delay, processing delay, queuing delay, and round-trip time (RTT). Optimizing the latency of the application has been a goal of many previous works, some of them based on autonomous networking in charge of detecting degradation of QoS and performing adaptations, e.g. using reinforcement learning in \cite{9907047} \cite{9771590}.

Thus, measuring packet latency, either hop-by-hop or end-to-end, is critical to assess whether or not a given network is capable of supporting latency-demanding services. In this regard, Inband Telemetry strategies implemented on P4 switches (PINT) can provide the tools to monitor individual links and end-to-end paths~\cite{Cugini22}. %However, telemetry may add latency to packets as they traverse PINT-based links, since the packets need to be tagged with such control information, which introduces extra delay to those packets~\cite{eucnc_jah}. %carrying monitoring information. 
%Alternative sentence:
However, telemetry consumes bandwidth resources due to its extra-headers, introduces additional forwarding operations, and it may overwhelm telemetry collectors. 

Taking only a few monitoring samples of highly variable metrics (like latency) may lead to wrong conclusions, but on the other hand, too many telemetry samples may burden the network excessively. In this sense, Cochran's formula~\cite{cochran} for infinite populations allows the appropriate dimensioning of sampling strategies to achieve high accuracy with moderate error. The formula relates the necessary number of monitoring samples $n_0$ with the error $e$ at estimating the percentage $p=1-q$ of packets below some delay threshold $D_{th}$, for some confidence level $Z$:
\begin{equation}\label{eq_cochran}
    n_{0} = \frac{Z^{2}\cdot p\cdot q}{e^2},
\end{equation}

\begin{figure*}[!htbp]
   \centering
    \includegraphics[width=1\linewidth]{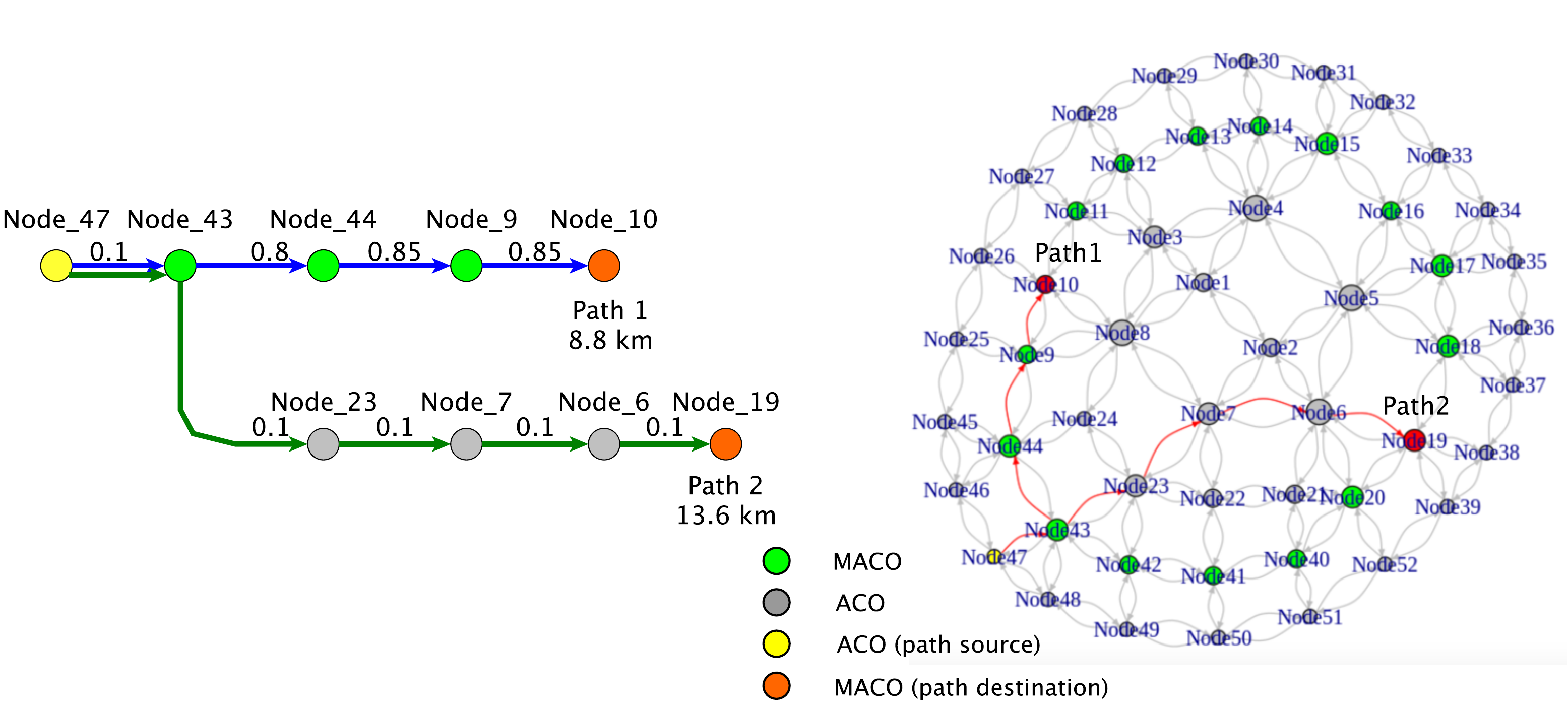}
    \caption{Two-path scenario in the Milano-net \cite{MilanNet} topology with $18.4$~Km of network diameter.}
    \label{fig:threelinkscenario}
\end{figure*}

As an example, consider we want to estimate the percentage of packets below some latency threshold say $D_{th}=82~\mu s$ between source Node\_47 and destination nodes Node\_10 and Node\_19 (path1 and path2), for the Milano-net~\cite{MilanNet} topology shown in Fig.~\ref{fig:threelinkscenario}, with 95\% confidence (i.e. $Z=1.96$ for 95\% confidence level). %\footnote{There is a variant of this formula for finite populations, but samples will be taken from a big enough number of packets so an infinite population can be assumed. In any case, this would be a worst-case scenario.}. 

In this scenario, if only $n_0=10$ latency measurements are collected, then the error is $e=0.3099$ or $\pm$30.99\% (from eq.~\ref{eq_cochran}). This error value reduces to $\pm$13.86\% for $n_0=50$ measurements and $\pm$4.90\% for $n_0=400$ measurement samples. Thus, Cochran's formula allows us to identify how many samples are needed for estimating delay percentiles accurately with bounded error.

As shown in Fig.~\ref{fig:threelinkscenario}, the two source-destination paths have different distances (8.8~Km and 13.6~Km) and have to traverse multiple hops with various link loads. Queuing delays are simulated as classical M/M/1 queues, where the total delay per link $T_i$ (including queuing and transmission) follows an exponential distribution with mean $E(T_i) = E(X)\frac{1}{1-\rho_i}$, that is, its Cummulative Distribution Function (CDF) follows:

\begin{equation}
F_{T_i}(t) = 1-e^{-\frac{t}{E(T_i)}}
\end{equation}
Here, $E(X)=1~\mu s$ is the average packet service time at 10~Gb/s and $\rho_i \in (0,1)$ is the load of the $i$-th link in the network. Propagation delay is computed following the classical $5~\mu s/Km$ for silica fiber. 

\begin{figure}[!htbp]
\centering
\includegraphics[width=0.75\columnwidth]{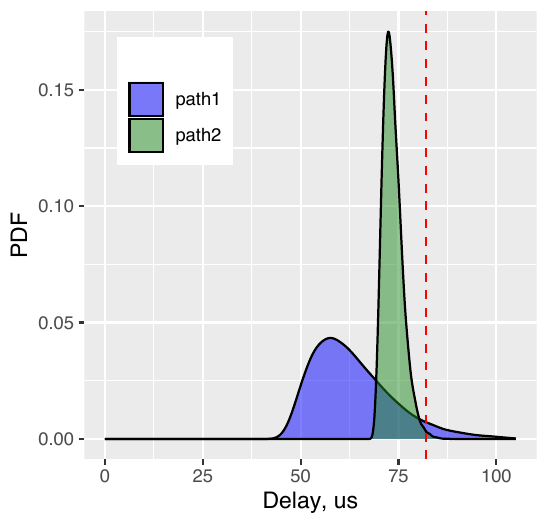}
\includegraphics[width=0.75\columnwidth]{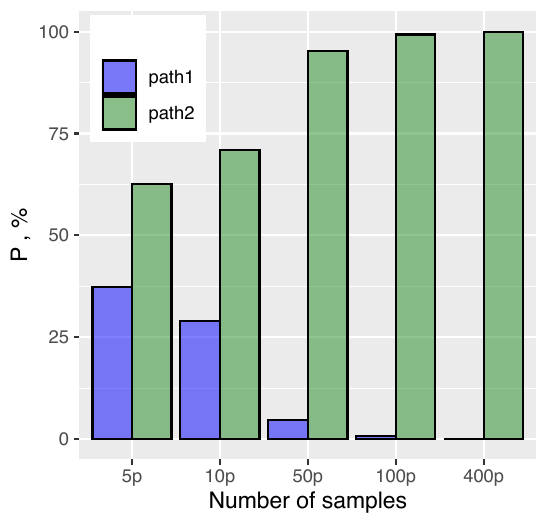}
\caption{The simulated distribution of packet delays for two paths with the threshold of 82 $\mu s$ (top), the percentage of times where each path is chosen as best in 10.000 experiments for the threshold of 82 $\mu s$ (bottom).}
\label{fig:3paths_pdf_barplot}
\end{figure}

Fig.~\ref{fig:3paths_pdf_barplot} (top) shows the Probability Density Function (PDF) of the simulated total delay (including propagation, queuing and transmission) for the two paths, along with the $82~\mu s$ delay threshold (red dashed). As shown, although the blue path has a shorter average delay ($63~\mu s$ average for blue and $73~\mu s$ for green), the green path has a smaller delay tail and is expected to meet the $82~\mu s$ requirement better than path 1 (93.7\% for blue vs 99.5\% for green below the threshold). 

%Path1 has 63.47 us average delay, 93.73% below 82 us. Path2 - 73.55 us and 99.526%

In this scenario, the control plane is assumed to receive telemetry information as a number of $n_0$ latency measurements and must decide which of the two paths can better guarantee the $82~\mu s$ delay threshold for a certain latency-critical application. Fig.~\ref{fig:3paths_pdf_barplot} (bottom) shows the estimated percentage of times where path1 is selected (blue) vs path2 (green) in the cases where 5, 10, 50, 100, and 400 latency measurements $n_0$ are collected. 

As depicted, if only $n_0=5$ measurements are collected (high error according to Cochran's formula), there is a high chance (in fact 37\% of the times) that the control plane wrongly chooses path1. When the number of latency measurements is 100 or above, the control plane mostly decides that path2 is more suitable than path1 at guaranteed delays below $82~\mu s$. Hence, if poor telemetry information is provided to the control plane, this may make wrong decisions very often.

\section{Full simulation set}

Let us consider again the network topology of Fig.~\ref{fig:threelinkscenario} (right) for Milano-net \cite{MilanNet} whose network diameter is 18.4 $Km$. In this scenario, we are interested in evaluating whether or not each source Access Central Office (ACO) can reach any data center available in the MAN, represented as green Metro-Aggregation Central Offices (MACOs) in less than $82~\mu s$ one-way delay. The total number of source-destination pairs evaluated is 35 ACOs x 17 MACOs, i.e. 595 paths. %To achieve this, we undertake the estimation of each path from each source node of 42 Local Central Offices (LCOs) to 10 nodes hosting data centers within the Milano-net, which comprises 10 central offices with associated data centers or National Central Offices (NCOs).

\begin{figure}[!htbp]
\centering
\includegraphics[width=1\columnwidth]{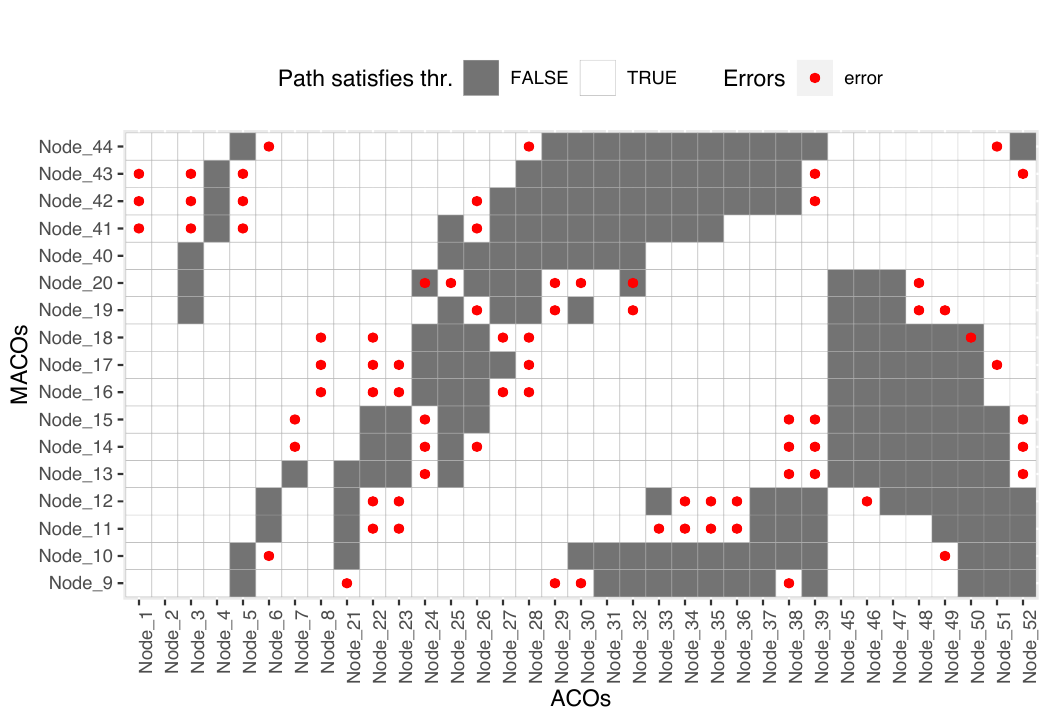}
\includegraphics[width=1\columnwidth]{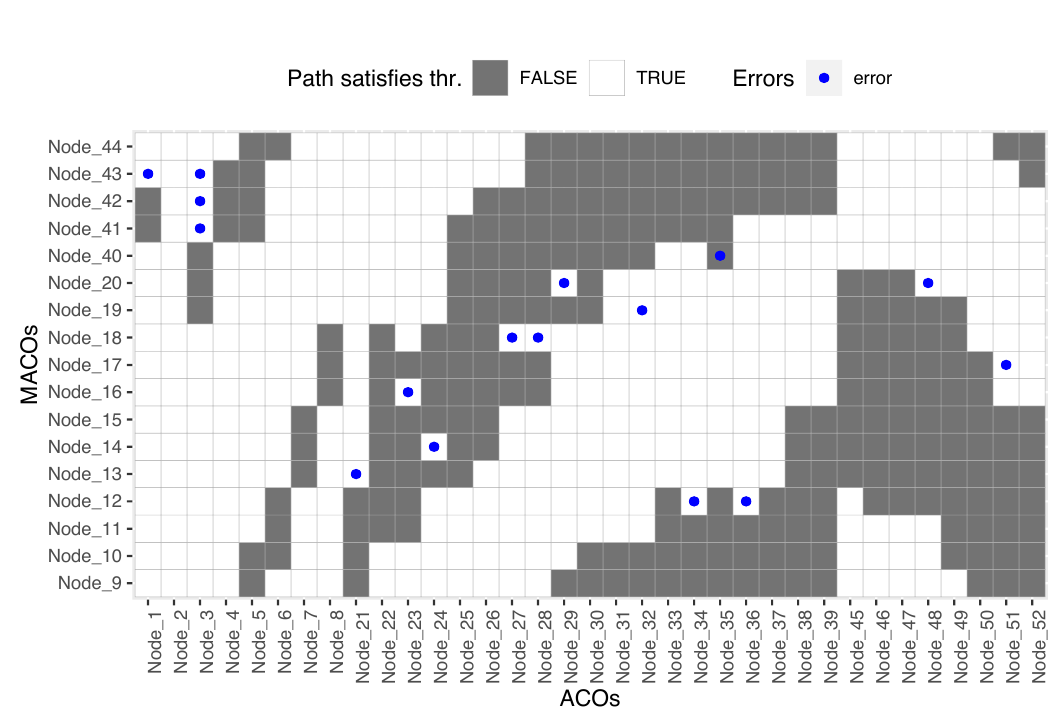}
\caption{Heatmaps of source-destination pairs below $82~\mu s$: (top) with 5p and (bottom) with 100p measurements.}
\label{fig:milano_sim_results}
\end{figure}

The heatmaps of Fig.~\ref{fig:milano_sim_results} illustrate whether or not each source-destination pair in the Milano topology is below $82~\mu s$, white for true, grey for false along with the errors as red/blue dots. In other words, for a path to fulfill the latency requirements, the delay of 99\% of sampled packets within that path must be below the specified threshold, $82~\mu s$ in the example. Fig.~\ref{fig:milano_sim_results} (top) presents telemetry-provided information using only a sample size of 5p latency measurements.

Fig.~\ref{fig:milano_sim_results} (bottom) provides the same information using 100 packets. %  Highlighted paths indicate instances where the results obtained from 5 or 100 packets deviate from those obtained with the entire population. 
In grey, we observe source-destination pairs not fulfilling the $82~\mu s$ requirement. The red dots indicate False Positive (FP) or False Negative (FN) errors due to the 5p sampling strategies. %When using 100p samples, the number of errors.
In particular, in the case of only 5p measurement samples, we find 73 FP and 3 FN errors. For 100p, these values reduce to 15 FP and only 1 FN. Finally, for 2500p (not depicted in Fig.~\ref{fig:milano_sim_results}), only 1 FP and 1 FN are obtained.

\section{Conclusion}

This article shows the importance of collecting accurate and sufficient telemetry information, especially for random metrics with high variability like latency. Accurate telemetry is critical as a step prior to Zero-Touch Networking scenarios, since insufficient measurements may lead to wrong decisions. We provide a methodology based on Cochran's formula to find the number of measurements needed for accurate delay estimates. Future work will evaluate such telemetry strategy in a P4 Docker container over SONiC with real traffic traces. 

The code used in the simulations is open-source and available in Github for the interested reading willing to reproduce the results and further extend the methodology with new scenarios and applications~\cite{koneva}. %\footnote{https://github.com/NataliaBlueCloud/Inband\_telemetry\_design.git}.

% conference papers do not normally have an appendix

% use section* for acknowledgment
\section*{Acknowledgment}

The authors would like to acknowledge the support of Spanish projects ITACA (PDC2022-133888-I00) and 6G-INTEGRATION-3 (TSI-063000-2021-127) and EU project SEASON (grant no. 101096120).

% trigger a \newpage just before the given reference
% number - used to balance the columns on the last page
% adjust value as needed - may need to be readjusted if
% the document is modified later
%\IEEEtriggeratref{8}
% The "triggered" command can be changed if desired:
%\IEEEtriggercmd{\enlargethispage{-5in}}

% references section

% can use a bibliography generated by BibTeX as a .bbl file
% BibTeX documentation can be easily obtained at:
% http://mirror.ctan.org/biblio/bibtex/contrib/doc/
% The IEEEtran BibTeX style support page is at:
% http://www.michaelshell.org/tex/ieeetran/bibtex/
%\bibliographystyle{IEEEtran}
% argument is your BibTeX string definitions and bibliography database(s)
%\bibliography{IEEEabrv,../bib/paper}
%
% <OR> manually copy in the resultant .bbl file
% set second argument of \begin to the number of references
% (used to reserve space for the reference number labels box)

% that's all folks
\end{document}